\documentclass[aps,twocolumn,pre]{revtex4-1}

\usepackage{graphicx}
\usepackage{amsmath}
\usepackage{amssymb}
\usepackage{amsbsy}
\usepackage{bm}
\usepackage{setspace}
\usepackage{amsfonts}
\usepackage{textcomp}
\usepackage{gensymb} 

\begin{document}

\title{A physical perspective on cytoplasmic streaming (invited)}
\author{Raymond E. Goldstein$^{1}$ and Jan-Willem van de Meent$^{2}$}
\affiliation{$^{1}$Department of Applied Mathematics and Theoretical
Physics, Centre for Mathematical Sciences, University of Cambridge, Wilberforce Road, Cambridge CB3 0WA, UK\\
$^{2}$Department of Engineering Science, University of Oxford, Parks Road, Oxford OX1 3PJ, UK}

\date{\today}

\begin{abstract}Organisms show a remarkable range of sizes, yet the dimensions of a single
cell rarely exceed $100$ $\mu$m.  While the physical and biological origins of this constraint remain poorly 
understood, exceptions to this rule give valuable insights.   A well-known counterexample is the 
aquatic plant \emph{Chara}, whose cells can
exceed $10$ cm in length and $1$ mm in diameter. Two spiraling bands of molecular motors at the cell
periphery drive the cellular fluid up and down at speeds up to $100$ $\mu$m/s, motion that has been
hypothesized to mitigate the slowness of metabolite transport on these scales and to aid in homeostasis. 
This is the most organized instance of a broad class of continuous motions known as 
``cytoplasmic streaming", found in a wide range of  eukaryotic organisms - algae, plants, 
amoebae, nematodes, and flies - often in unusually large cells. In this overview of the
physics of this phenomenon, we examine the interplay between streaming, transport and cell size, and discuss
the possible role of self-organization phenomena in establishing the observed patterns of streaming.
\end{abstract}

\maketitle

\section{Transport and Cell Size in Biology}

Relative to the remarkable variation of sizes exhibited by living organisms, the dimensions of a
typical cell are surprisingly similar across species. In nearly all plant-like multicellulars the
constituent cells measure $10-100$ $\mu$m, with animal cells falling into a similar if slightly smaller
range and single-cellular prokaryotes extending down to $1$ $\mu$m. Thus outside of a few notable
exceptions the overwhelming majority of organisms have cell sizes in the range $1-100$ $\mu$m. A very
basic question about the fundamentals of cellular biology is:  What underlying mechanism has
determined the evolution of this relatively well-conserved length scale? 

The fact that very few cells are larger than $100$ $\mu$m suggests that this size reflects physical
constraints, such as the diffusive range over which two metabolites can reliably interact. The
outliers showing larger cell sizes provide key counterexamples, prompting the question of how these
species have compensated for problems associated with an increasing cell size, and what utility this
design provides in an evolutionary context. In some cases large sizes are found in cells with a
highly specialized role. The nerve fibres in our body can reach lengths of a meter, and the
\emph{giant squid axon} can have a diameter up to $1$ mm, improving the propagation speed of the
action potentials which ultimately facilitates a faster escape response. In other cases,
single-cellular organisms have evolved to a large size and complexity. Examples of this are the
protozoan \emph{Paramecium}, which can reach $350$ $\mu$m, and the trumpet-shaped \emph{Stentor} 
($2$ mm).  Arguably even more developed is the alga \emph{Acetabularia}, a single-cellular organism that 
grows into a plant-like stalk that can be as long as $10$ cm. Finally there are the characean algae, a
family of plant-like weeds whose segmented stems are built up out of cylindrical \emph{internodal
cells} of $200-1000$ $\mu$m diameter and lengths that can exceed $10$ cm.
 
Cells in these examples have the common feature of forms of active internal
transport, driven by the movement of \emph{molecular} motors along intracellular filaments that make
up the \emph{cytoskeleton}. In small cells, this motion enables movement of organelles and vesicles.
In larger cells it leads to a continuous circulation of the cellular fluid, known
as \emph{cytoplasmic streaming} or \emph{cyclosis}. Streaming is found in many types
of larger eukaryotic cells, particularly in plants \cite{allen_arbb_1978}.
Patterns of circulation take on a variety of geometric forms and can be very steady. 

Cytoplasmic streaming has long been conjectured to aid in overcoming the slowness of diffusion on
long length scales, but its precise role in enhancing metabolic rates is yet to be elucidated. This
article provides an overview of the current state of knowledge surrounding this issue. Over the last
two decades, our understanding of the molecular make-up and spatial organization of the cellular
environment has increased dramatically. We begin with an overview of how spatial aspects of
intracellular transport factor in to our understanding of metabolic regulation. This is followed by
an explanation of the range of topologies of circulation found across species. The arguably best
studied instance of this phenomenon is the \emph{rotational} streaming found in the long cylindrical
cells of the characean algae, whose metabolic context is covered in detail. We conclude with a
discussion of those studies that have investigated the role of circulation in intracellular
transport and provide an overview of key questions in furthering our understanding of how streaming
enables enhanced diffusive transport.

\subsection{Homeostasis and Targeting of Macromolecules}

Cellular life requires precise control of metabolic pathways and biosynthesis. Two central
challenges in internal regulation are the maintenance of \emph{homeostasis} and the control of
\emph{trafficking}. In metabolic pathways, turnover rates can vary by several orders of magnitude.
The dynamic equilibrium of a biological system's properties in the face of fluctuating levels of
throughput is generally termed homeostasis \cite{cannon_book_1932}. The concept of homeostatic
control necessarily encompasses a range of regulatory mechanisms, whose function is often specific
to the type of  organism considered. In mammals, key homeostatic requirements are the regulation of
body temperature and the stabilization of oxygen pathways and cellular ATP levels under varying
levels of muscle activity \cite{hochachka_pnas_1999}.  In plants, homeostatic control is associated
with turgor pressure and cytosolic concentrations of inorganic ions such as calcium and
phosphate \cite{boller_arpp_1986}, as well as cellular building blocks such as amino acids.
 
In addition to homeostatic regulation, which predominantly involves concentrations of small
molecules,
intracellular transport of proteins, lipids and polysaccharides requires a high degree of targeting 
\cite{schwartz_ari_1990, chrispeels_arpppmb_1991, staehelin_arpp_1995, juergens_arcdb_2004}.
Eukaryotic cells possess an array of organelles, each containing a distinct set of proteins that
allow it to perform its biochemical activities. The translation of nearly all cellular proteins
takes place in the cytoplasm, after which they must be targeted to one of more than 30
compartments in the cell \cite{schwartz_ari_1990}. The mechanisms for targeting form a
central area of inquiry in cell biology and have been studied intensively in recent years. We
now know that each protein contains a targeting domain in its amino acid sequence, which interacts
with receptors on the organelle to which it is to be transported. Additionally, many macromolecules
are transported between organelles inside lipid vesicles, with a differentiated coating, as well as
a SNARE marker protein to identify its contents \cite{juergens_arcdb_2004}.

Two organelles that play a central role in protein synthesis are the endoplasmic reticulum (ER) and
the Golgi apparatus. The ER is a network of continuous tubules that courses through the cytoplasm.
Comprising as many as 16 functionally distinct subdomains, it fulfills a multitude of roles in
cellular metabolism and in some species extends between cells through the channels known as
plasmodesmata \cite{staehelin_pj_1997}. The ER is the entry point for newly synthesized proteins into
the trafficking network of membrane-bound organelles known as the endomembrane system. Inside the
ER, polypeptides created by ribosomal translation are folded into soluble proteins by means of the
chaperone BiP \cite{juergens_arcdb_2004}. Typically, protein products from the ER are subsequently
extruded in vesicles and transported to the Golgi apparatus for further sorting and processing.
Storage proteins, for example those utilized as nutrients in seed formation, can also bypass the Golgi
apparatus to be stored in protein storage vacuoles (PSVs) by way of intermediate
compartments \cite{juergens_arcdb_2004}.

The plant Golgi apparatus plays a central role in protein processing and sorting, but also
synthesises large quantities of cell wall polysaccharides and glycolipids for inclusion in
the plasma membrane \cite{staehelin_arpp_1995, dupree_bba_1998, juergens_arcdb_2004}. Plant cells
can have tens to hundreds of Golgi stacks, which are often closely associated 
with the ER network and are known to localize to specific subcellular regions in cell types 
exhibiting localized cell wall growth \cite{dupree_bba_1998}.

\subsection{Diffusion in a Crowded Cytoplasm}

\begin{figure}[t]
\begin{center}
  \includegraphics[width=\columnwidth]{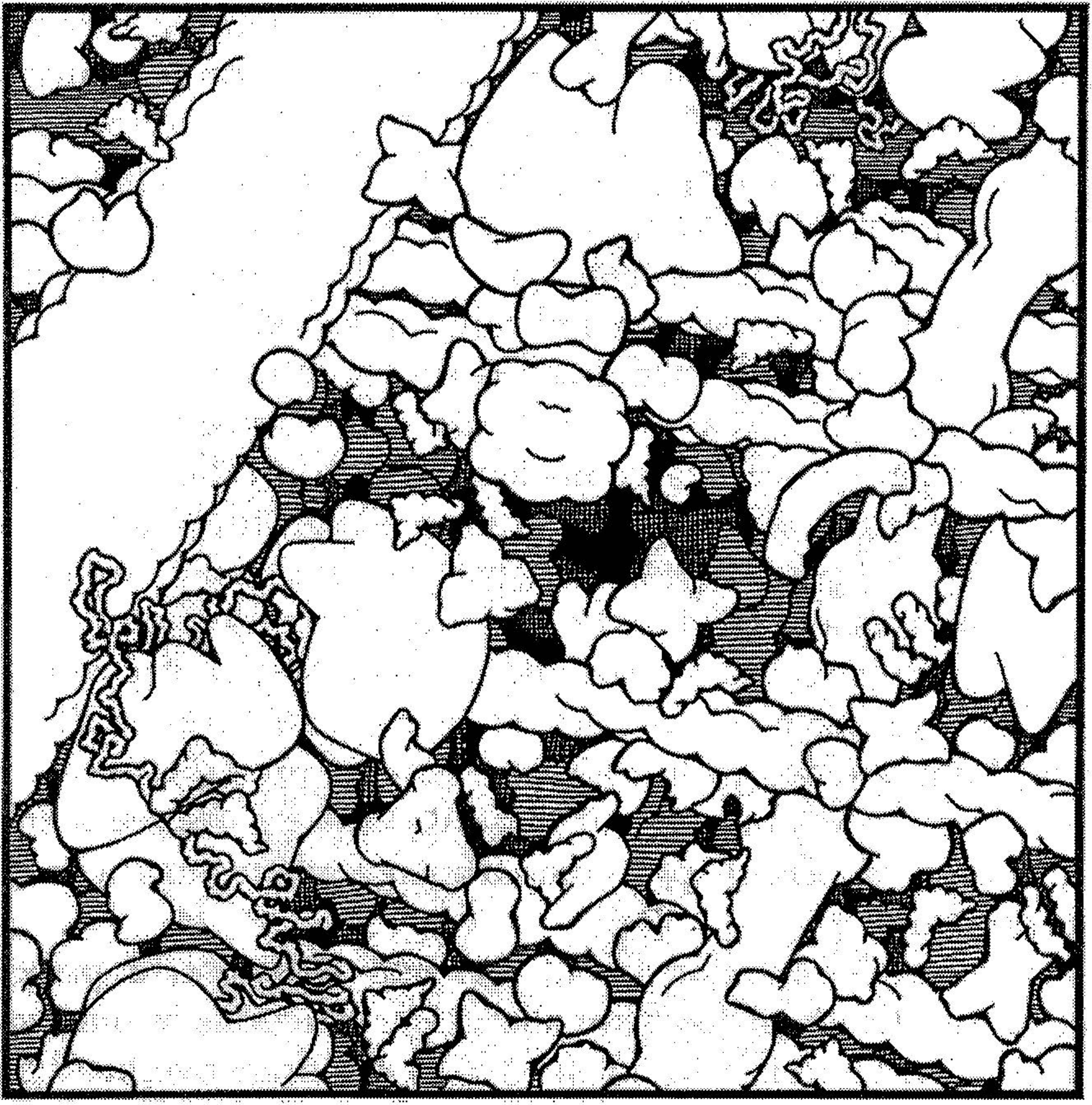}
  \caption{Molecular crowding in eukaryotic cytoplasm. Shown is an illustration of the contents of the yeast
  \emph{Saccharomyces cerevisiae}. Proteins, ribosomes with mRNA, microtubules, actin filaments and 
  intermediate filaments are all drawn to scale and at physiological concentrations. 
(Adapted with permission from \cite{goodsell_book_2009}.)} 
\label{fig:goodsell}
\end{center}
\end{figure}

In formulating an understanding of trafficking and homeostatic control, it is becoming increasingly
apparent that many transport processes in cells require explanation \cite{hochachka_pnas_1999,
agutter_jhb_2000, agutter_be_2000}.   In the past, a simplified picture of a cell as a watery bag of
enzymes has often been invoked for lack of a more detailed model. This metaphor has given way to the
notion of a highly crowded and high structured cytoplasm, requiring a spatial (and chemical)
understanding of metabolic pathways. As demonstrated in a comprehensive review by
Luby-Phelps \cite{luby-phelps_irc_2000}, a number of assumptions valid for chemical reactions in an
aqueous environment break down in the cytoplasm. 

\emph{Reaction volumes are small.} While biochemical analysis typically neglects finite-volume
effects, key molecules \emph{in vivo} are often only present in limited numbers. Physiologically relevant
concentrations tend to lie in the nM-mM range. A single molecule inside a bacterium has a nominal
concentration of $10$ nM, while eukaryotic cells may possess perhaps only $1000$ copies of a molecular
species \cite{alberts_book_2002}.  Thus, the number concentration of subcellular compartments may in
many cases be more informative than the nominal molar concentration.

\emph{The cytoplasm is a crowded environment.} In a cellular environment macromolecules typically
occupy $20-30$\% of the total volume \cite{ellis_tbs_2001}, leading to a range of phenomena known
together as \emph{macromolecular crowding} \cite{luby-phelps_irc_2000, ellis_tbs_2001,
zimmerman_arbbs_1993, minton_cobt_1997}.  Figure \ref{fig:goodsell} shows an iconic visualization of
the cytoplasmic environment in baker's yeast \cite{goodsell_book_2009}.  Cellular components such as
proteins and ribosomes are all shown at correct scale and density, along with cytoskeletal elements.
While $70-80$\% of the volume is water, it is clear that diffusive movement will be constrained in a
manner that will depend strongly on particle size.

This intuitive notion is born out by experimental evidence. Measurements of mobility of water in
living tissue by NMR and quasi-elastic neutron scattering (QENS) show a reduction of the order of
$50$\% of both rotational and translational diffusion \cite{luby-phelps_irc_2000, clegg_ajpricp_1984,
rorschach_sm_1991, cameron_cbi_1997}. Self-diffusion of larger macromolecules shows a strong
reduction in mobility that depends on the hydrodynamic radius \cite{luby-phelps_irc_2000,
verkman_tbs_2002}, declining from $40$\% to $5$\% of the aqueous diffusion rate for particle sizes
increasing from $2$ nm to $45$ nm \cite{luby-phelps_irc_2000, luby-phelps_jcb_1986,
luby-phelps_pnas_1987, popov_jns_1992, arrio-dupont_bpj_1996}. More recently, several studies have
presented evidence of \emph{anomalous diffusion} in the cytoplasm: particle trajectories show a
mean squared displacement that is subdiffusive, i.e. $\langle r^2(t) \rangle \sim t^{\alpha}$ with
$\alpha < 1$ \cite{wachsmuth_jmb_2000, weiss_bpj_2004, banks_bpj_2005, golding_prl_2006}. For larger
vesicles the diffusion rates are of order 10$^{-11}$ cm$^2$/s, implying that vesicular movement will
be insignificant in the absence of active transport mechanisms \cite{luby-phelps_irc_2000}.

\emph{The cytoplasm is highly structured and compartmentalized.} It is increasingly recognized that the
cytoplasm is not a homogeneous environment. Macromolecular crowding effects may well result in
separation of the cytoplasmic volume into distinct microphases, as evidenced from phenomena such as
caged diffusion of microinjected beads and the size-dependent partitioning of inert 
tracers \cite{luby-phelps_irc_2000}.

\subsection{Motor-driven transport along the cytoskeleton}

In the spatial organization of a cell, the \emph{cytoskeleton} plays a central role. This meshwork
of \emph{actin}, intermediate filaments and \emph{microtubules} acts as a backbone for directed
transport. Myosin molecular motors can bind to cytoplasmic structures and transport them along actin
filaments by a `walking' motion that consumes ATP. Microtubules have kinesin and dynein motors that
perform similar tasks. The filaments that make up the cytoskeleton are polar, in the sense that
molecular motors move in a well-defined direction along their tracks. Whereas the various myosins
walk along actin in directions that depend on their type, kinesins walk towards the plus end of
microtubules, which typically results in  transport from the cell centre towards the periphery. 
Dyneins travel in the opposite direction. The topology of the cytoskeleton thus defines the
direction of motor-assisted transport.

The forms of active transport observed depend greatly on the organism and cell type. In animal
cells, which tend to be small compared to their plant counterparts, organelles are relatively
stationary and motors are primarily implicated in the transport of vesicles \cite{schwartz_ari_1990},
whose diffusion is negligibly slow \cite{luby-phelps_irc_2000}.   A review of organelle movements in
plant cells by Williamson \cite{williamson_arpppmb_1993} shows that movement of nearly all major
organelles has been observed. Golgi stacks move over ER strands in an actomyosin dependent 
manner \cite{boevink_tpj_2002, juergens_arcdb_2004}, while the ER itself is also thought to bind to myosin
motors \cite{kachar_jcb_1988, williamson_arpppmb_1993, staehelin_pj_1997, du_jcs_2004}. There is also
evidence for actomyosin driven movement of mitochondria \cite{vangestel_jebo_2002}. Chloroplasts are
known to move along with streaming, as well as move themselves to positions determined by light and
cell division planes, though the details of chloroplast-cytoskeleton contact are often less well
understood \cite{williamson_arpppmb_1993}.

\subsection{Mixing in the Cytoplasm}\label{sec:mixing}

As Purcell famously outlined in his paper ``Life at low Reynolds number"
\cite{purcell_ajp_1977}, fluid flows at the cellular scale are dominated by viscosity, where
our intuitions shaped by life in a high $Re$ world do not apply. The Reynolds number
$Re = UL/\nu$ is the dimensionless ratio of inertial and viscous forces, containing a typical
velocity $U$, a system size $L$, and the kinematic viscosity $\nu$ of the fluid. The 
properties of low Reynolds flows have important implications for mixing behavior. 
A drop of milk in a glass of tea spreads in a turbulent cloud when stirred. As Taylor 
illustrated in his 1967 film
``Low Reynolds number flows" \cite{taylor_film_1966}, this `mixing' is reversible in a low $Re$
flow. If a blob of dye is injected in a cylinder filled with a very viscous fluid, it
apparently dissolves over a few turns of a stirrer, but if the stirrer's motion is reversed the 
blob returns to its original shape, blurred only slightly by diffusion. 

In considering diffusion of a solute of concentration $c$ and diffusion constant $D$ in the
presence of a flow field ${\bf u}$ we may rescale lengths by $L$ and time by $L^2/D$, 
to obtain the dimensionless advection-diffusion equation
\begin{align} \label{intro:advdiffpe}
 c_t  + Pe ({\bf u} \cdot \nabla) c &=  \nabla^2 c,
\end{align}
where the P{\'e}clet number $Pe=UL/D$ characterizes the relative strength of advection to diffusion
in a form analogous to the Reynolds number. Small molecules in aqueous solutions have $D\sim
10^{-5}~\mathrm{cm^2/s} =  1000~\mathrm{\mu m^2/s}$. A $1$ $\mu$m bacterium, swimming at $10$ $\mu$m/s,
would have $Pe \sim 0.01$. The smallest organisms therefore not only live in a world where the
Reynolds number is essentially zero, they also live in a world where diffusive fluxes tend to
outpace advective fluxes. Put more plainly: for a bacterium, it is equally efficient to sit in one
place and let food arrive by diffusion as it is to swim to get it. Locomotion does not serve to
increase uptake directly as much as it allows migration to areas richer in
nutrients \cite{purcell_ajp_1977}.

As the system size becomes larger, \mbox{$Pe = UL/D$} and \mbox{$Re  = UL/\nu$} both increase,
since they share a dependence on $UL$, but since $D$ is typically three (or more) orders of magnitude
smaller than $\nu$, the P{\'e}clet number becomes significant long before the Reynolds number 
does.  By definition, advection competes with diffusion when $Pe
\sim 1$, and this provides us with an estimate of the dimensions at which flow could potentially aid
cellular metabolism. For a typical length of $10$ $\mu$m and a flow rate of $1$ $\mu$m/s, structures with a
diffusion constant lower than 10 $\mathrm{\mu m^2 / s}$ will start to be affected by internal
circulation. This number roughly corresponds to the diffusion rate of vesicles in the
cytoplasm \cite{luby-phelps_irc_2000}, indicating that active transport of vesicles may be worthwhile
even in small cells. For small molecules, on the other hand, diffusion constants of order $1000$
$\mathrm{\mu m^2 / s}$ imply system sizes closer to $100$ $\mu$m for
transport rates in the range of $10$ $\mu$m/s.
So we see that active transport phenomena, which have typical velocities of $1-10$ $\mu$m/s, could well
be significant in aiding intracellular transport. However, in the absence of turbulent mixing, the
precise mechanism by which transport is facilitated in the presence of a flow does need
qualification. Some of the mechanisms by which flows may aid molecular transport at low Reynolds
numbers are outlined below.

\subsection{Chaotic Flow Fields.} ~One mechanism often discussed for increasing diffusion and
reaction rates involves chaotic advective fields, in which trajectories of neighboring points diverge
exponentially over time \cite{aref_jfm_1984, aref_pf_2002}. Mixing enhancement of this type has been
studied extensively for microfluidic and lab-on-a-chip systems \cite{squiresquake}.  Examples
include the herringbone micromixer, in which a chaotic circulation pattern is induced by a spatially
alternating pattern of grooves in the bottom of a channel \cite{stroock_sci_2002}, and mixing in a
droplet undergoing periodic deformations induced by electrowetting \cite{mugele_apl_2006}. A
pictorial way of describing these effects is that the surface described by the interface between two
solutions is repeatedly stretched and folded into itself, much like the swirl of a Danish pastry, resulting in a
drastic reduction of the length over which molecules have to diffuse. 

\subsection{Taylor Dispersion.} ~Another well-known mechanism of dispersion at low Reynolds
numbers is \emph{Taylor dispersion}, an enhancement of effective diffusivity due to shear that was
originally described by G.I. Taylor in the context of pressure driven pipe
flows \cite{taylor_prsa_1953, taylor_prsa_1954, taylor_ppsb_1954}.  A layer of solute carried along by
such a flow is deformed into a parabolic sheet by the velocity profile. This results in outward
diffusion at the leading edge of the profile, and inward diffusion at the tail. After a time long
enough for diffusion to smooth out the deformed sheet radially, it converges to a self-similar
profile that travels with the mean flow speed and disperses with an effective diffusion constant
\begin{equation} \label{eq:taylor-dispersion}
 D_{\rm eff} \simeq D \left(1 + \frac{1}{192} Pe^2 \right) ~,
\end{equation}
which can greatly exceed the bare value $D$.
In applying such concepts to subcellular mixing it must be recognized that results
obtained in long channels or many repeated drop deformations do not necessarily translate to
the intermediate length- and time scales found in single cells.  As the 
time scale for diffusion across a channel of height $h$ is $h^2/D$, whereas the distance traveled 
along the channel in this time is $U h^2 /D = Pe h$, we should therefore expect the 
Taylor dispersion to become observable only when the aspect ratio $L/h > Pe > \sqrt{192}$. 

\subsection{High P{\'e}clet Enhancement of Reactive Fluxes.} ~An example of how advection 
can affect nutrient uptake was found in theoretical studies of the \emph{Volvocales} \cite{short_pnas_2006,
solari_pnas_2006}. These algae take the form of spherical colonies that swim through their
environment by the action of flagellated cells on their surface. The flow at the leading edge of
the colony compresses a solute boundary layer closer to the surface of the cell, while a depleted
plume is formed at the trailing edge. This effect produces a boundary layer whose size scales as 
$Pe^{-1/2}$, resulting in an enhanced concentration gradient and thereby an
increased flux into the organism \cite{short_pnas_2006}.
This flux enhancement is but one instance of a more general phenomenon. For \emph{life at high P\'eclet
numbers}, the boundary layers around metabolically active objects will shrink with increasing
strength of the flow field, thereby facilitating increased fluxes into the object.  Pickard has
recently applied these ideas to analysis of moving objects in the cytoplasm \cite{pickard_jtb_2006}.
Approximating vesicles and organelles as spherical objects he showed that the advective flux scales
as $Pe^2$ for $Pe \lesssim 1$, and crosses over to a $Pe^{1/3}$ scaling for $Pe \gtrsim 10$. At
$Pe = 1$, fluxes should be expected to be increased by roughly 100\% over the purely diffusive value.

\begin{figure*}[t]
  \includegraphics[width=0.95\textwidth]{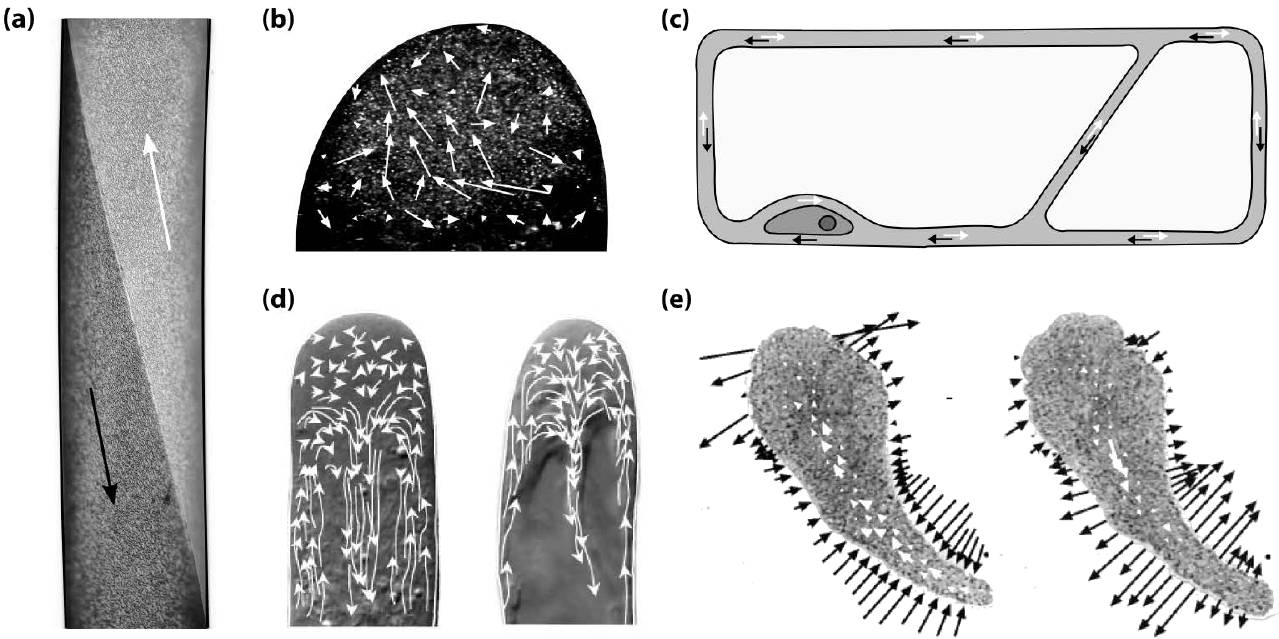}
  \caption{\label{fig:streaming_topologies} Topologies of cytoplasmic streaming. 
  (a) Rotational streaming in internodal cells of \emph{Chara corallina}. 
  (b) A developing oocyte of \emph{Drosophila} exhibits a correlated random flow field 
  with typical flow rates of a few to tens of nm/s. Schematic of the velocity field, extracted by particle
image 
  velocimetry. 
  (c) Circulation streaming in the periphery and transvacuolar strand of epidermal cells, e.g. as
  found in the root of \emph{Medicago truncatula} \cite{sieberer_pro_2000}. 
  (d) Reverse-fountain streaming in \emph{Lilium longiflorum} (left) and 
  \emph{Nicotiana tabacum} (right). (Image modified from \cite{hepler_arcdb_2001}.)  
  (e) Periodic shuttle streaming in plasmodium fragments of \emph{Physarum polycephalum}. 
  (Adapted from \cite{matsumoto_bpj_2008}.)}
\end{figure*}

In summary, as a result of advances in fluorescence labeling techniques and biochemical knowledge
over the past decades, we are now arriving at an understanding of the eukaryotic cytoplasm as a
dynamic environment that is both crowded and highly structured, possessing tight homeostatic control
mechanisms as well as a motor-driven infrastructure for precise routing of proteins towards their
destination. It is increasingly becoming clear that quantitative understanding into trafficking and
homeostatic control in this environment will require formulation of a spatially resolved picture of
signalling and reaction rates along metabolic pathways. 

Active transport processes mediated by molecular motors may play a key role in extending the spatial
range over which metabolites can be reliably expected to interact, and it is in this context that
the role of continuous forms of circulation observed in some larger cells is of interest. While the
precise effect of advection on diffusive transport depends critically on the type of geometry
studied, various mechanisms for enhancement of molecular transport can be identified for flows where
the P\'eclet number is sufficiently large. The remainder of this review provides an overview of
what is known about cytoplasmic streaming and its effect on intracellular transport. We pay
particular attention to what is perhaps the best studied instance of this phenomenon, the
\emph{rotational} streaming that occurs in the giant \emph{internodal} cells of the
characean algae, but also indicate emerging areas of interest in other organisms.

\section{Cytoplasmic Streaming}

Cytoplasmic streaming has been known for more than two centuries, with the first observations
attributed to Bo\-na\-ven\-tu\-ra Corti in 1774 \cite{corti_book_1774}.  It occurs in a range of cell
types and a variety of organisms including amoebae, protozoa, fungi and slime moulds, 
and is also found during oogenesis in the fruit fly \cite{ganguly} and 
embryogenesis in the nematode \emph{Caenorhabditis elegens} \cite{streaming_elegans},
though it is most
common in plants and their highest genetic predecessors, the \emph{characean
algae} \cite{kamiya_book_1959, kamiya_inproc_1962, kamiya_arpp_1981, allen_arbb_1978,
allen_arbb_1978b}. 
A large body of research over the last five decades has established that
streaming is in most instances driven by an actomyosin system \cite{shimmen_cocb_2004,
shimmen_jpr_2007}, but there are also examples of microtubule-based organelle
movement \cite{nebenfuehr_pp_1999, mizukami_pcp_1981, mineyuki_pro_1986}. Though the mechanics of
cytoplasmic streaming have been studied extensively, there is relatively little insight into its
biological function. A number of authors, most notably Pickard, have suggested that streaming may
enhance metabolic rates in large cells, where diffusive time scales become prohibitively 
large \cite{pickard_pro_1974, pickard_pce_2003, hochachka_pnas_1999}.

Streaming has customarily been classified into two major groups. The term \emph{amoeboid} streaming
denotes cytoplasmic motion that induces changes in cell form, the best known of which is
\emph{shuttle streaming} found in slime moulds \cite{kamiya_book_1959, kamiya_arpp_1981,
allen_arbb_1978b}. Non-amoeboid streaming is generally divided into five classes based on visually
apparent phenomenology, as originally identified by Kamiya \cite{kamiya_book_1959, allen_arbb_1978,
pickard_pce_2003}. The best studied by far is the \emph{rotational streaming} found in characeans,
which will be reviewed in detail in the next section. Other forms of streaming commonly identified
are \emph{saltation}, \emph{circulation}, \emph{fountain streaming} and \emph{multi-striate
streaming}, briefly described below. 

\emph{Shuttle Streaming} (fig \ref{fig:streaming_topologies}e) is a periodic flow found in the
\emph{plasmodium} of slime moulds. The plasmodium is a single-celled aggregate of protoplasm
containing many nuclei. As the organism searches for food it forms a network of veins, known as
\emph{pseudopodia}, that self-optimizes in connecting to food sources, and can find the shortest
routes through mazes \cite{nakagaki_nat_2000, nakagaki_prslb_2004, tero_sci_2010}.  Rhythmic
back-and-forth streaming of the cytoplasm takes place inside these filaments, driven by contraction
waves in the actomyosin network. The velocities of this pressure-driven motion, which reverses every
$2-3$ minutes, can reach $1350$ $\mu$m/s \cite{kamiya_book_1959}.
%add citations\cite{kamiya_arpp_1981,
%kamiya_pro_1959, kamiya_pro_1950, kessler_book_1982, teplov_bs_1991, smith_bpj_1992,
%takamatsu_prl_2000, matsumoto_bpj_2008}

\emph{Random Streaming} is a spatially correlated yet apparently
unordered motion of the cytoplasm. The best-known
example of this is found in various developmental stages of
{\it Drosophila} \cite{mahajan-miklos_db_1994,
palacios_dev_2002, serbus_dev_2005, zimyanin_cel_2008}.  There, within an oocyte several hundred microns across, 
kinesins moving along a dense network of microtubules producing streaming flows that vary from one oocyte to another,
and over time within a given oocyte, and consists of swirls and eddies with a correlation length of some $20$ $\mu$m 
\cite{ganguly}.   The nature of transport on such a disordered network is under active study \cite{KhucTrong}.

\emph{Saltation}, also known as \emph{agitation}, is the most widespread form of cytoplasmic
movement \cite{kamiya_book_1959, allen_arbb_1978}, characterized by apparently random jumps of
cytoplasmic particles over distances as large as $100$ $\mu$m, much greater than those
corresponding to thermal fluctuations. Microtubules are possibly implicated in this form of
streaming, since very active saltation is observed near spindles, within centrosomes and adjacent to
microtubular organelles \cite{allen_arbb_1978}.

\emph{Circulation} (fig \ref{fig:streaming_topologies}c) takes the form of movement along the cell
wall and strands of cytoplasm transecting the vacuole. Circulation patterns are typically stable on
the time scale of minutes, and evolve as transvacuolar strands move and branch. Both unidirectional
and bidirectional movement are known and velocities can be as large as $40$ $\mu$m/s. Cells exhibiting
circulation include hair cells in various plants such as \emph{Urtica} (stinging nettle), and
parenchymal (\emph{i.e.}~bulk) cells in \emph{Allium} (the onion genus), as well as the leaf cells of the
water plant \emph{Elodea}.\cite{kamiya_book_1959, allen_arbb_1978}.

\emph{Fountain streaming} (fig \ref{fig:streaming_topologies}d) is circulation in which the
cytoplasm moves along a central axis, flowing back in the opposite direction along the cell wall.
\emph{Reverse-fountain streaming} exhibits an inward motion along the central axis and is the more
common of the two \cite{kamiya_inproc_1962, allen_arbb_1978}. In some cell types it is a
developmental stage towards rotational streaming. True fountain streaming is typically found in
root hairs and pollen tubes of various plants \cite{vondassow_jcb_1994, miller_pro_1996,
miller_jebo_1997, tominaga_pro_1997, sieberer_pro_2000, vidali_pro_2001, hepler_arcdb_2001,
derksen_spr_2002, lovy-wheeler_cyt_2007}.

\emph{Multi-striate streaming} is found in the fungus \emph{Phycomyces} and in the marine alga
\emph{Acetabularia}. The cells of \emph{Acetabularia} have a cylindrical stalk several centimeters
in length, containing a large vacuole separated by a tonoplast from a thin layer of
\emph{cytoplasm}, at the periphery. Streaming occurs in both directions along channels separated by
stationary cytoplasm \cite{allen_arbb_1978}.  Some forms of circulation could arguably be seen as
multistriate streaming, such in the marine alga \emph{Caulerpa}, where streaming forms $100$ $\mu$m wide
bands wherein files of chloroplasts stream bidirectionally in a circadian
rhythm \cite{allen_arbb_1978}.

\section{The characean algae}

\begin{figure*}[t]
\includegraphics[width=\textwidth]{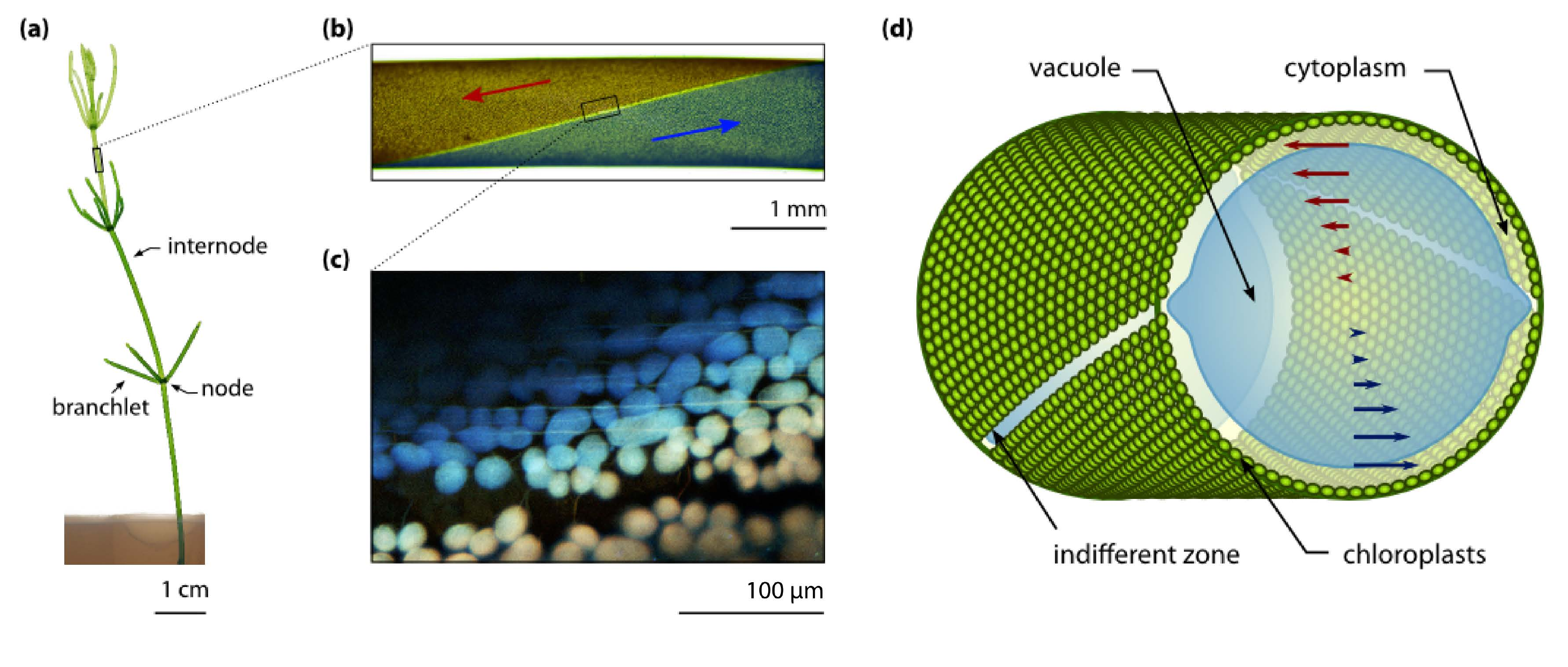}
\caption{\label{fig:chara_anatomy} Rotational streaming in the characean algae. (a) A shoot
of \emph{Chara corallina} anchored in agar. Single-celled internodes connect nodal complexes where a
whorl of 6 branchlets is formed. (b) Cytoplasmic streaming takes place along two domains
shaped as spiraling bands. (c) This circulation is driven by the motion of myosin molecular
motors along bundled actin filaments. This image shows a merged stack of confocal slices, with the
colors denoting the focal position. Actin bundles can be observed below chloroplast rows at the
surface of the cell. (Image courtesy S. Ganguly.) (d) The motion of myosin at the periphery
entrains the outer layer of cytoplasm, which is of order $10$ $\mu$m in thickness. The two moving bands are
separated by a neutral line visible as a row of missing chloroplasts. The motion at the wall induces
a shear flow in the central vacuole of the cell. } 
\end{figure*}

One of the most studied examples of cyclosis is the rotational streaming in giant cylindrical cells
of the characean algae, or Charales (fig \ref{fig:chara_anatomy}). Colloquially known as
\emph{stoneworts} after the lime deposits on their surface, these plant-like species are found in
dense meadows on the bottom of lakes and ponds. \emph{Chara} cells have been studied since the early
days of microscopy \cite{corti_book_1774} and the species are now recognized as the closest living
relatives of land plants \cite{kranz_jme_1995, karol_sci_2001}.  This high degree of similarity, and
the robustness of internodal cells under manipulation, has made them a model organism in a wide
range of plant physiology research, including membrane transport and 
electrophysiology \cite{shimmen_csf_1994}, turgor-driven cell wall 
expansion \cite{kamiya_pro_1963, green_sci_1967,
green_pp_1971, zhu_pp_1992, proseus_jebo_2000, proseus_jebo_2006b, proseus_jebo_2007} and
cytoskeletal organization \cite{wasteneys_pla_1992, wasteneys_ba_1993, wasteneys_cmc_1993,
wasteneys_pro_1996, kropf_pro_1997, braun_pla_1998, foissner_pla_1999}, calcification and carbon
fixation \cite{lucas_jebo_1975}, intercellular transport through plasmodesmata \cite{ding_pcp_1989,
kikuyama_pcp_1992, franceschi_pla_1994, lucas_cocb_1995, blackman_pj_1998}, and even lake
ecology \cite{vandenberg_ab_1998, chen_em_2002, kufel_ab_2002, siong_jeq_2006, rodrigo_ab_2007}.

Found in fresh and brackish waters, the characean algae have the appearance of plants, growing in
thin segmented shoots that sprout whorls of branchlets every few centimeters
(\ref{fig:chara_anatomy}a). Their most studied part is the internode (fig \ref{fig:chara_anatomy}b),
a single cylindrical cell with a diameter up to 1 mm and a length that can exceed 10 cm. Like most
plant cells, it is enclosed by a cellulose-containing cell wall, lined by a layer of \emph{cytoplasm} $\sim 10$
$\mu$m in thickness. A membrane known as the \emph{tonoplast} separates the cytoplasm from the
central vacuole that occupies the bulk of the cell. This vacuole fulfils a multitude of
metabolic roles, acting as storage compartment for sugars, polysaccharides and organic acids,
sequestering toxins such as heavy metals, and functioning as a buffering reservoir that helps to
maintain ionic and pH homeostasis in the cytoplasm \cite{taiz_jeb_1992}.  Additionally, the vacuole
holds a $0.13$ M concentration of salts \cite{tazawa_pcp_1964}, producing a turgor 
pressure equivalent to $5$ bar that lends the cell its rigidity. 

The characean cytoplasm contains many structures that are common to higher plants, including the
Golgi apparatus, Endoplasmic Reticulum (ER) and cytoskeletal filaments. \emph{Chloroplasts}, the
organelles responsible for photosynthesis, are packed into helical rows that spiral along the cell
surface (fig \ref{fig:chara_anatomy}d). The chloroplasts are surrounded by a stagnant layer of
fluid, the \emph{ectoplasm}, that houses most of the \emph{mitochondria} in the
cell \cite{foissner_pro_2004}. On the inside of the chloroplast rows, bundled actin filaments (fig
\ref{fig:chara_anatomy}c) act as tracks for myosins that drag structures within the cell
\cite{kachar_sci_1985, kachar_jcb_1988}, and thereby entrain the inner part of the cytoplasm, the
\emph{endoplasm}. With streaming rates as high as $100$ $\mu$m/s, the myosin XI found in {\it Chara} is
the fastest known in existence \cite{shimmen_cocb_2004}.  As a result of a reversed polarity of the
actin filaments, the cytoplasm is organized into two bands flowing in opposite directions. These bands
spiral around each other, producing a ``barber-pole" velocity at the cell periphery (fig
\ref{fig:chara_anatomy}c-d). The two interface lines between these bands are known as \emph{neutral
lines} or \emph{indifferent zones}. They are marked by the absence of chloroplasts (fig
\ref{fig:chara_anatomy}d) and are visible as two light lines crossing the cell surface. 

The internodes connect to the nodal structures by means of channels known as
\emph{plasmodesmata}. These channels are bridges of cytoplasm between cells. The plasmodesmata in
Charales are similar to those found in higher plants, but may have evolved separately \cite{franceschi_pla_1994}.
Most notably, in higher plants the endoplasmic reticulum is known to
extend through plasmodesmata, but characean algae lack this feature. The generally accepted size
limit for these channels is 800-1000 Da \cite{lucas_np_1993} for diffusive transport, though
molecules as large as 45 kDa have been shown to move between cells on longer time 
scales \cite{kikuyama_pcp_1992}.

\subsection{Rate of Streaming and Velocity Profile}

\begin{figure*}
   \includegraphics[width=\textwidth]{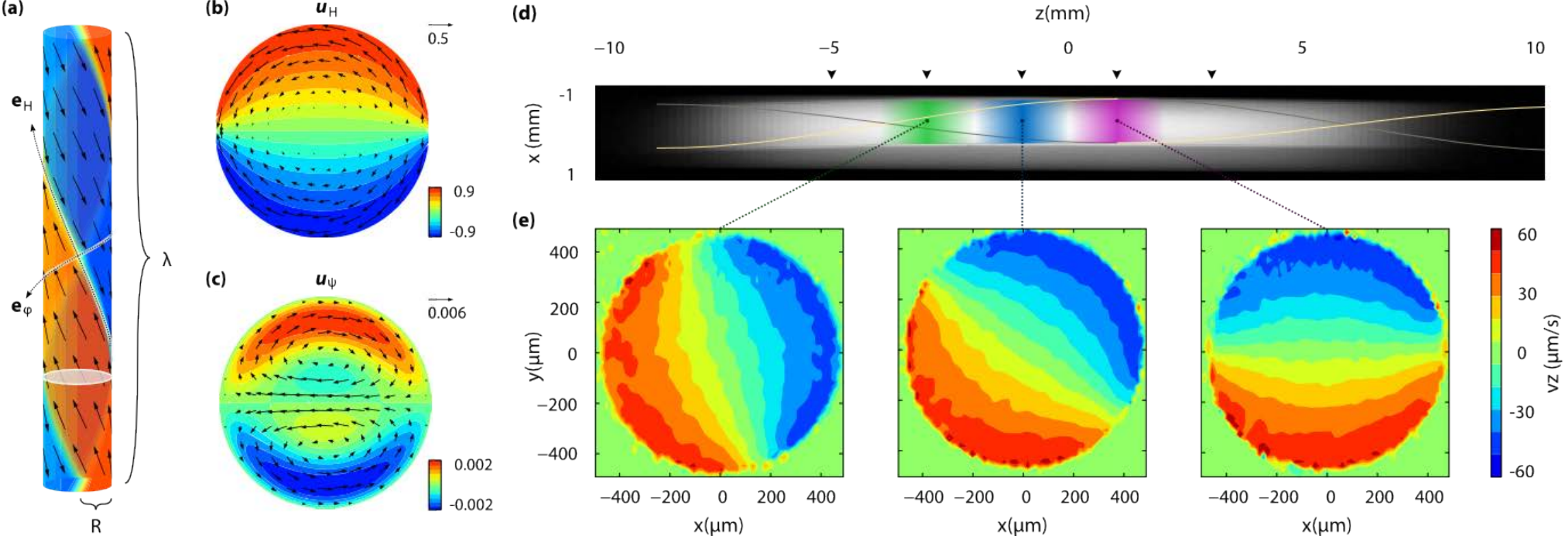}
   \caption{\label{fig:chara_flow_profiles} Hydrodynamic prediction and MRV measurements of vacuolar
  flow. (a) The internodal flow has a helical symmetry: an invariance under a
  translation along the longitudinal axis combined with a rotation. The two axes that naturally
  follow from this symmetry are the vector $\vec{e}_H$, which points along the bands, and the vector
  $\vec{e}_\varphi$, which is orthogonal to the bands. (b) Theoretically predicted flow field
  along the $\vec{e}_H$ component, showing the vacuolar shear profile. (c) The $\vec{e}_r$ and
  $\vec{e}_\varphi$ components reveal a small secondary circulation along the center of the cell.
  (d) MRI scan of an internode placed in a glass tube, with spiraling lines indicating
  positions of the neutral line, and coloured bands showing the domains used for velocity
  measurements. (e) Velocity profiles measured at each of the domains show excellent agreement
  with the theoretical profile. (Figures modified from \cite{goldstein_pnas_2008, vandemeent_prl_2008,vandemeent_jfm_2010}.)}
\end{figure*}

The symmetry of characean internodes makes them amenable to a relatively straightforward
hydrodynamic description. Internodal cells have a very large aspect ratio, typically exceeding 30, 
so, for positions sufficiently far from the endpoints, the flow is well approximated as that inside
an infinite cylinder, with the cytoplasmic bands effectively imposing a value for the velocity at
the boundaries (fig \ref{fig:chara_flow_profiles}a). The Reynolds number $Re = UR/\nu$ is at most
$0.05$, so the full hydrodynamic equations reduce to the linear \emph{Stokes flow}, in which
\begin{align}
  \label{eq:stokes_flow}
  \eta \nabla^2 {\bf u}  &= \nabla p ~,    &
  \nabla \cdot {\bf u} &= 0 ~.
\end{align}
Pickard solved the simplest case of this flow problem for \emph{Chara}, assuming straight,
non-twisting indifferent zones lying along the z-axis \cite{pickard_cjb_1972}. In this case, solving
Stokes flow reduces to solving Laplace's equation on a circle and the solution of the form $u_z \sim
\sum W^n r^n sin(n \theta)$ is obtained readily by separation of variables. Assuming step boundary
conditions (velocity $\pm U$ on the two halves of the circle) allows a closed-form solution
\begin{equation}\label{eq:pickard_closedfrm}
  u_z ~=~ \frac{2 U}{\pi} \arctan \left( \frac{2 (r/R) \sin \theta}{1-(r/R)^2} \right)~.
\end{equation}
A general solution for the case of twisting helical bands of wavelength $\lambda$ can be obtained as
Fourier-Bessel series \cite{goldstein_pnas_2008, vandemeent_prl_2008}. A decomposition along the axis
of symmetry (fig \ref{fig:chara_flow_profiles}a) shows that the flow field possesses two components,
a downstream term that reduces to the profile (\ref{eq:pickard_closedfrm}) in the limit of infinite
helical pitch, and a small-amplitude circulation between the neutral lines. The amplitude of the
second term vanishes as $\lambda\to 0$ and $\lambda\to\infty$, reaching a maximum for $\lambda/R\sim 3$. During
development of the internodes, the helical bands twist and subsequently untwist, attaining a minimum
in the helical pitch roughly at the stage of maximal growth \cite{green_ajb_1954}. It is therefore
possible that this secondary circulation, though modest in amplitude, aids homeostasis by enhancing
vacuolar mixing \cite{goldstein_pnas_2008, vandemeent_prl_2008}.

A series of increasingly sophisticated experiments, employing first visual
tracking \cite{kamiya_bmt_1956, pickard_cjb_1972} and later laser Doppler
spectroscopy \cite{mustacich_bpj_1977}, has shown that the approximation of a simple shear flow is in
good agreement with the \emph{in vivo} observations. Recently,  full 2-dimensional measurements of
the velocity field were obtained using magnetic resonance velocimetry (fig
\ref{fig:chara_flow_profiles}c-d) which confirm that the vacuolar flow is virtually
indistinguishable from its hydrodynamic prediction \cite{vandemeent_jfm_2010}, with direct shear
transmission by the tonoplast.  The role of a lipid membrane in allowing such direct shear transmission has
recently been investigated theoretically \cite{WoodhouseJFM} and experimentally \cite{HonerkampSmith}.

\subsection{Cell Development}

In order to extend towards sunlight, plant cells expand their volume 10- to 20-fold over the course
of development \cite{green_pp_1971, taiz_jeb_1992}. The majority of this volume is occupied by the
central vacuole, where high concentrations of salts are sequestered, resulting in large outward
osmotic pressures that maintain the rigidity, or \emph{turgor}, of the cell. Perhaps one of the most
important functions of these vacuolar balloons is to serve as an energetically favorable means of
expanding the volume of the organism. While there is a considerable cost associated with the
transport against electrochemical gradients required to maintain the solute concentration in the
vacuole, it is much cheaper to expand cell size by water uptake than by protein 
synthesis \cite{raven_np_1981, taiz_jeb_1992}. With their exceptionally long internodal cells, whose vacuoles
occupy about 95\% of the cellular volume, characean algae represent an extreme case of this growth
strategy. Since these species grow as weeds in low-nutrient conditions and can outcompete higher
plants for lake dominance \cite{coops_ab_2002}, the large cell size in these organisms may be
partially understood as a result of competition for light in low nutrient conditions.

Growth in characean algae occurs by repeated division of a single cell at the plant tip, the
\emph{apical cone} \cite{green_jbbc_1958}.  During shoot development this cell divides every few days.
The rate of this division, sometimes called the \emph{plastochron}, can be quite regular.  After
division, the newly formed cell divides once more into an upper \emph{pro-node} and an internode.
The internode undergoes a remarkable degree of expansion, increasing in volume by five orders of
magnitude during $2-3$ weeks \cite{green_jbbc_1958}. The pro-node undergoes a number of subsequent
divisions to form the nodal structure. In \emph{Chara corallina}, there is an initial division into $6$
cells which connect to the ascending and descending streams in an specific manner \cite{shepherd_pce_1992b}.
In later stages these cells undergo further divisions to form a complex
multicellular structure, though the initial symmetry is still reflected in the fact that \emph{C.
corallina} typically forms a whorl of 6 branchlets. The reproductive structures are exclusively
formed on the part of the node connecting to the descending stream \cite{shepherd_pce_1992b}.

\subsection{Nutrient Uptake and Intercellular Transport} \label{sec:lit_intercel_trans}

One of the most obvious roles of cytoplasmic streaming in characean metabolism is to enhance
the rate of transport between cells, thereby facilitating the translocation of nutrients from
regions of uptake to regions of growth. Because of their relative anatomical simplicity, the
Charales have been used for a great number of uptake and transport studies using radioisotopes
to track salts
\cite{bostrom_jebo_1975, williams_jebo_1975, zawadzki_jebo_1986, zawadzki_jebo_1986b} and nutrients
such as inorganic carbon \cite{williams_jebo_1975, zawadzki_jebo_1986, zawadzki_jebo_1986b,
trebacz_jebo_1988, ding_pcp_1989}, nitrogen, and phosphorus \cite{littlefield_pp_1965, box_pce_1986,
box_pce_1987, vermeer_ab_2003}.

Like most aquatic macrophytes, the Charales are able to take up nutrients from the surrounding
water, in contrast to land plants where uptake predominantly takes place in the roots. While the
species can successfully be cultured in flasks without needing to be anchored in a layer of 
soil \cite{forsberg_pp_1965}, there is evidence that a significant proportion of uptake may take place in
the root-like \emph{rhizoidal} cells that anchor the plants, particularly for rate-limiting
nutrients such as phosphorus and nitrogen which may be more abundant by orders of magnitude in the
interstitial water of the sediment \cite{littlefield_pp_1965, box_pce_1986, box_pce_1987,
vermeer_ab_2003}.

Dissolved inorganic carbon (DIC) is taken up directly from the water surrounding the shoot.
Low-weight carbohydrates are formed within $1$ hour after uptake of $^{14}$C-DIC and pass unaltered
through the plasmodesmata to the neighboring cell \cite{trebacz_jebo_1988}. The bulk of the
photoassimilates reside in the endoplasmic layer on the inside of the chloroplasts, but particularly
in branchlet cells some fraction is sequestered in the vacuole \cite{ding_jebo_1991b}. The branchlet
cells are more photosynthetically active than the internodes, and transport from the
branchlets to internodes can be five times higher than the reverse \cite{ding_jebo_1991b,
ding_jebo_1991}.

Many studies have investigated the relationship between streaming and intercellular transport.
Measurements with a tandem pair of internodes show that the transport rate correlates with the
streaming rate over the $25$\% variation in magnitude observed in a collection of 
samples \cite{zawadzki_jebo_1986}. This correlation is stronger in summer when the rate of transport is
higher \cite{shepherd_pce_1992,shepherd_pce_1992b,zawadzki_jebo_1986}.
Treatment with cytochalasin B shows a roughly proportional reduction of the transport rate with the
streaming velocity \cite{bostrom_jebo_1976}. Treatment of either cell in the pair suffices to lower
the transport rate; a similar response is found when altering the streaming velocity by lowering the
temperature \cite{ding_pcp_1989}. Some early studies found indications of a small component of
propagation faster than the streaming velocity \cite{williams_jebo_1975, dale_jebo_1983}, but this
finding was not confirmed by later whole-shoot experiments where $^{14}$C is fed to the rhizoids,
showing that carbon is transported upwards at roughly the rate of streaming while $^{32}$P
transport is somewhat slower \cite{box_jebo_1984}.

\subsection{Alkaline Band Formation and Carbon Fixation}\label{sec:lit_carbon_alkaline}

The lime deposits from which the species get the name \emph{stoneworts} arise from their alkaline
habitat. Carbon fixation associated with photosynthesis naturally leads to production of OH$^-$, as
can be seen in the pH dependence of the chemical equilibrium of DIC in water,
\begin{equation}
  \mathrm{CO_{2} + H_{2}O  \rightleftarrows H^{+} + HCO_{3}^{-}} ~.
\end{equation}
At pH $5.5$ roughly $90$\% of the dissolved carbon is found in the form CO$_2$, but as the pH
increases this balance shifts towards HCO$_{3}^{-}$. At pH $8.5$, which is fairly typical for
environments inhabited by characeans, $99\%$ of the inorganic carbon is in HCO$_3^-$. A proton is
therefore required to produce CO$_2$ from bicarbonate, which inevitably leaves OH$^-$ as a side
product. This excess OH$^{-}$ is excreted at the surface, inducing a rise in pH. In
characeans OH$^{-}$ is localized in periodically spaced alkaline bands alternated by acidic regions
where H$^{+}$ efflux occurs \cite{tazawa_arpp_1987}. Calcium carbonate (CaCO$_3$) tends to
precipitate on the alkaline regions forming the band-shaped encrustations characteristic of
stoneworts. This precipitation yields a proton,
\begin{equation}
  \mathrm{Ca^{2+} + HCO_{3}^{-} \rightarrow CaCO_{3} + H^{+}} ~,
\end{equation}
which may then be used to produce CO$_2$ from bicarbonate. It has been suggested that the H$^{+}$
efflux serves to facilitate photosynthesis by raising the concentration of free 
CO$_2$ \cite{plieth_pp_1994}. Evidence that this precipitation enhances carbon fixation is found in
$^{14}$C studies that show a $1:1$ correspondence between the rate of CaCO$_3$ precipitation and the
rate of carbon fixation in slightly alkaline environments of pH $8-9$ \cite{mcconnaughey_bb_1991,
mcconnaughey_lo_1991, mcconnaughey_cjb_1998}.

Further support for the notion the pH bands enhance photosynthesis is found in the work by Mimura 
\emph{et al.} \cite{mimura_pce_1993}, who investigated the effect of the 
H$^+$-ATPase on $^{14}$C fixation. They observed that carbon fixation was strongly reduced both when 
the ATP in the cytoplasm
was depleted by perfusion with hexo\-kinase and 2-deoxyglucose and when the H$^+$-ATPase was
inhibited by introduction of vanadate. Stimulation of the H$^+$-ATPase by pyruvate kinase and
phosphoenolpyruvate (PEP) resulted in increased carbon fixation. The same effects were observed at
both pH 5.5 and pH 8.5, indicating that the H$^+$-ATPase plays an essential role in assimilation of
both CO$_2$ and HCO$_3^-$.

Evidence that the formation of alkaline bands directly influences the rate of photosynthesis is also
found in \emph{fluorometry} studies. When a sample is excited with light, the
absorbed light can be used for photosynthesis, emitted as fluorescence or dissipated as heat. In
practice the degree of heat dissipation can often be taken as constant, so the fluorescence signal
represents an inverse measure of the rate of photosynthetic activity. Such measurements show that
the external pH correlates directly with the rate of photosynthetic activity, with the highest rates
observed in acidic regions \cite{bulychev_pla_2003, bulychev_bcm_2005}.

Alkaline band formation appears inseparably linked to cytoplasmic streaming. The magnitude of pH
variations is greater for the upwards streaming band than for the downwards streaming
band \cite{babourina_jebo_2004}, and inhibition of cytoplasmic streaming with cyto\-chalasin B prevents
formation of alkaline bands \cite{tazawa_arpp_1987, lucas_jmeb_1977}. The membrane of internodal 
cells can undergo an \emph{action potential}, a transient depolarization similar to that observed
in nerve cells, although the travel speed of about $1$ cm/s is much slower. During an action potential, 
streaming is momentarily halted and the alkaline bands disappear. As streaming recovers, the 
alkaline bands tend to reform at the same position on the cell's surface and the recurrence was
increased with the concentration of Ca$^{2+}$ in the surrounding medium \cite{eremin_pps_2007}.

Although models for the formation of alkaline bands have been presented \cite{toko_jtb_1985}, the 
specific role of streaming in the generation of alkaline bands has not been widely discussed in 
the literature. A function of streaming thus far overlooked may well be to drive these bands, 
thereby aiding carbon uptake from the environment and enhancing photosynthetic rates.

\subsection{Driving Mechanics and Cytoplasmic Rheology} \label{sec:flow_rheol}

Various studies have investigated the driving mechanism and rheological aspects of cytoplasmic
streaming. A number of authors have presented theoretical models of transport along the cytoskeleton
with stochastic on-off dynamics \cite{dinh_bjbl_2006, snider_pnas_2004, maly_jtb_2002,
smith_bpj_2001}. The most recent work of this form applied to streaming is by Houtman 
\emph{et al.} \cite{houtman_epl_2007}, who describe streaming in transvacuolar strands with a 
two-dimensional model that
includes on/off kinetics and hydrodynamic interactions between particles through the Oseen tensor.
There is also a range of studies that focus on the driving force associated with streaming. Pickard
investigated the scaling of streaming with cell size \cite{pickard_pro_1974}, presenting an analysis
to support the notion that most dissipation occurs near the neutral lines. His measurements of the
streaming velocity as a function of cell size show the scaling of the maximum streaming 
speed $U \sim R^{1/2}$, which is consistent with the combined assumption of a dissipation rate scaling as 
$U^2$ and a driving power scaling as $R$.

Tazawa \& Kishimoto \cite{tazawa_pcp_1968} measured the motive force using perfusion experiments.  In this
technique, the cell is placed in an isotonic bath (\emph{i.e.} a solution of an osmolarity similar to that
of the vacuolar fluid), which allows the cell to survive amputation of its ends. If each of the
endpoints is then contained in a compartment separated from the rest of the bath, the contents of
the vacuole can be replaced by applying a slight pressure difference \cite{tazawa_pcp_1964,
tazawa_pcp_1964b}.  To measure the motive force, the pressure difference between the two reservoirs
is adjusted so that streaming is halted in one of the bands, implying that the shear force balances
the motive force in the cytoplasmic layer. The motive force obtained this way has a value in the
range of $14-20$ $\mu$N/cm$^2$, consistent with centrifuge microscope
measurements \cite{kamiya_pro_1958}. Of note here is that this force is independent of temperature.
Rates of cytoplasmic streaming show a widely documented linear (or even exponential) dependence on
temperature \cite{kamiya_book_1959, tazawa_pro_1968, pickard_pro_1974, mustacich_bpj_1976}, with
Pickard reporting an increase of $3.4$ $\mathrm{\mu m \:s^{-1} \: K^{-1}}$ \cite{pickard_pro_1974}.
Tazawa's results therefore suggests that this temperature dependence is a consequence of a change in
cytoplasmic viscosity. Tazawa \& Kishimoto also examined the effect of tonicity, finding that an increase in the
tonicity from $290$ mM to $586$ mM results in a lowered streaming rate as well as an increased
motive force. Decreasing the tonicity in the cytoplasm to $190$ mM resulted in a swelling of the
chloroplasts, causing a marked decrease in streaming rate and motive force, presumably because of a
resulting deformation in the actin bundles. After $5-20$ min, the chloroplasts regained their
normal shape and streaming recovered, though the motive force remained lowered.

Donaldson \cite{donaldson_pro_1972} performed measurements of the forward streaming velocity as a function of the applied
perfusion pressure gradient.  He assumed a driving force localized to a
layer of thickness $\epsilon$ and a power-law dependence $\tau = -\alpha (\partial u/ \partial y)^{1/n}$ for the
viscous stress. His results show a good correspondence for $n=3$ and  $\epsilon = 0.1$ $\mu$m. The
corresponding motive force, $F$ = $36$ $\mu$N cm$^{-2}$, is higher than the values found
by others \cite{tazawa_pcp_1968,kamiya_pro_1958}, which he concludes is the result of a 
systematic underestimation in those measurements due to the fact that the small velocities and thin
layers of movement near the stalling point are very difficult to observe.
Hayashi \cite{hayashi_jtb_1980} applied a similar analysis to experimentally measured velocity
profiles.  Like Donaldson he assumed a power-law rheology and found the
best-fit parameters for measurements of a protoplasm filled cell \cite{kamiya_bmt_1956}, as well as
two cases of plug flow of extracted cytoplasm \cite{kamiya_inproc_1965}. He found an exponent of $n =
1.4$ for the protoplasm-filled cell and exponents of $n = 1.3$ and $n = 1.7$ for two cases of
plug flow. 

Nothnagel and Webb \cite{nothnagel_jcb_1982} investigated various hydrodynamic models for driving
mechanics in the cytoplasm and concluded that the virtually shear-less profile observed in the
cytoplasm is best explained by assuming a meshwork throughout the cytoplasm that is pulled along at
the cell wall. The endoplasmic reticulum is argued to be a structure that could fulfill such a role
and there is structural evidence from electron micrographs to support this
hypothesis \cite{kachar_sci_1985}.  Recent work \cite{Wolff1} has examined the importance of hydrodynamic
slip at the cell wall.

\section{Role in Intracellular Transport}

Although a great deal of work has been published on the molecular basis and hydrodynamics of
streaming, relatively few authors venture into a discussion of its function. It has long been
suggested that streaming aids molecular transport in some way. Kamiya's 1959
review \cite{kamiya_book_1959} mentions that de Vries suggested this as early as
1885 \cite{devries_bz_1885}.  However, concrete hypotheses as to the mechanism by which streaming
accelerates metabolic rates have scarcely been put forward. Agutter \emph{et al.}~have argued that
diffusion is not capable of explaining many transport phenomena in cells \cite{agutter_jhb_2000,
agutter_be_2000}.  Similarly, Hochachka presents an argument that the degree of homeostasis along
ATP pathways cannot be explained other than by assuming forms of active
transport \cite{hochachka_pnas_1999}.

The highly symmetric topology of streaming in the characean algae would appear to have evolved at
considerable evolutionary cost, as further reflected in the fact that the myosin XI found in this
organism is the fastest known in existence. On the basis of what we know about the characean algae,
we see that streaming is implicated in a multitude of roles in cellular metabolism. It aids
transport between cells, and is therefore essential in supplying a steady flow of cellular building
blocks to newly formed cells at the tip of the shoot. It also appears important in maintaining the
alkaline bands that facilitate uptake of inorganic carbon from the surrounding water. However a key question
that remains largely unanswered is just what role streaming may play in eliminating the diffusional
bottlenecks that would seem to limit cell sizes in other organisms. Indeed streaming may help
homeostatic regulation during rapid cell volume expansion, but the precise mechanisms by which it 
does so remain an open area of investigation.
 
The most important contributions in terms of a quantified discussion of the effect of streaming on 
intracellular transport are without doubt from Pickard. He discussed scaling of the streaming velocity and 
diffusional time scales with cell size \cite{pickard_pro_1974}, as well as the interaction between the stagnant layer of
periplasm surrounding the chloroplast rows, and the moving layer of endoplasm. He points out the possibility that 
advection of a point-source may aid homeostasis by
smoothing out fluctuations in the concentration field. He also raises the notion that streaming as 
such does not necessarily have to confer a benefit to the cell if its real purpose is transport
of particles along the cytoskeleton. This second point is argued further in a later publication where he makes a
case for streaming being an accidental consequence for vesicular transport along the
cytoskeleton \cite{pickard_pce_2003}. As covered in section
1.4, he also discusses the role of boundary layer scaling in
enhancing exchange of molecular species between organelles
and their environment in advection-dominated flows \cite{pickard_jtb_2006}.

Our own work on the secondary circulation in the vacuole mirrors this last theme of life at high
P\'eclet numbers \cite{goldstein_pnas_2008, vandemeent_prl_2008}.  In the case of a cell with
helically twisted bands, the secondary circulation vortices in figure \ref{fig:chara_flow_profiles}c
produce an advection from center to wall and vice versa. In the case of a transient uptake into the
vacuole, this produces a boundary layer at the left side of the cross-section, where the flow is
directed outward, whilst diffusional pile-up is carried inward at the opposite neutral line.  The
boundary layer arising from this effect scales as $Pe^{-1/3}$.
This phenomenon suggests that the helicity of the bands, which is maximized at the moment of
greatest growth, may serve to aid cellular homeostasis by facilitating enhanced diffusive fluxes
into and out of the vacuole, allowing it to play a buffering role with respect to metabolic
processes in the cytoplasm. However, analysis of the eigenmodes of transient decay towards a
well-mixed vacuolar concentration shows that the enhancement of fluxes into and out of the vacuole
would become significant from $Pe_\psi \gtrsim 10$. This suggests that this type of effect would
primarily benefit transport of slowly diffusing structures, such as large macromolecules, whose
presence in characean vacuoles has currently not been established \cite{vandemeent_thesis_2010}.

\section{Streaming and Cell Size: Key Questions}

As emphasized in the mini-reviews in the above sections, the past two decades have led to 
remarkable progress in our understanding of the biochemical pathways central to cellular 
metabolism. In terms of the
dynamical behavior of these systems, we have seen that there are two central challenges in cellular
regulation. (1) Maintaining homeostatic levels of small molecules and ions
across varying levels of metabolic turnover and (2) The routing of proteins and other macromolecule
intermediates towards their intended targets.

A fundamental question from a biophysical point of view is what constraints have governed the
evolution of homeostatic control mechanisms and macromolecular targeting. It is clear that both these tasks 
are complex, and will lead to requirements that vary
between cell types and species. Yet while there are important differences, the functioning of the ER
and Golgi complexes shows a high degree of similarity across a wide range eukaryotic species. The
vesicle transport system in particular has been shown to be highly conserved, exhibiting essentially
the same mechanisms in organisms as diverse as mammals and yeast \cite{staehelin_arpp_1995}.

\begin{figure*}
   \includegraphics[width=\textwidth]{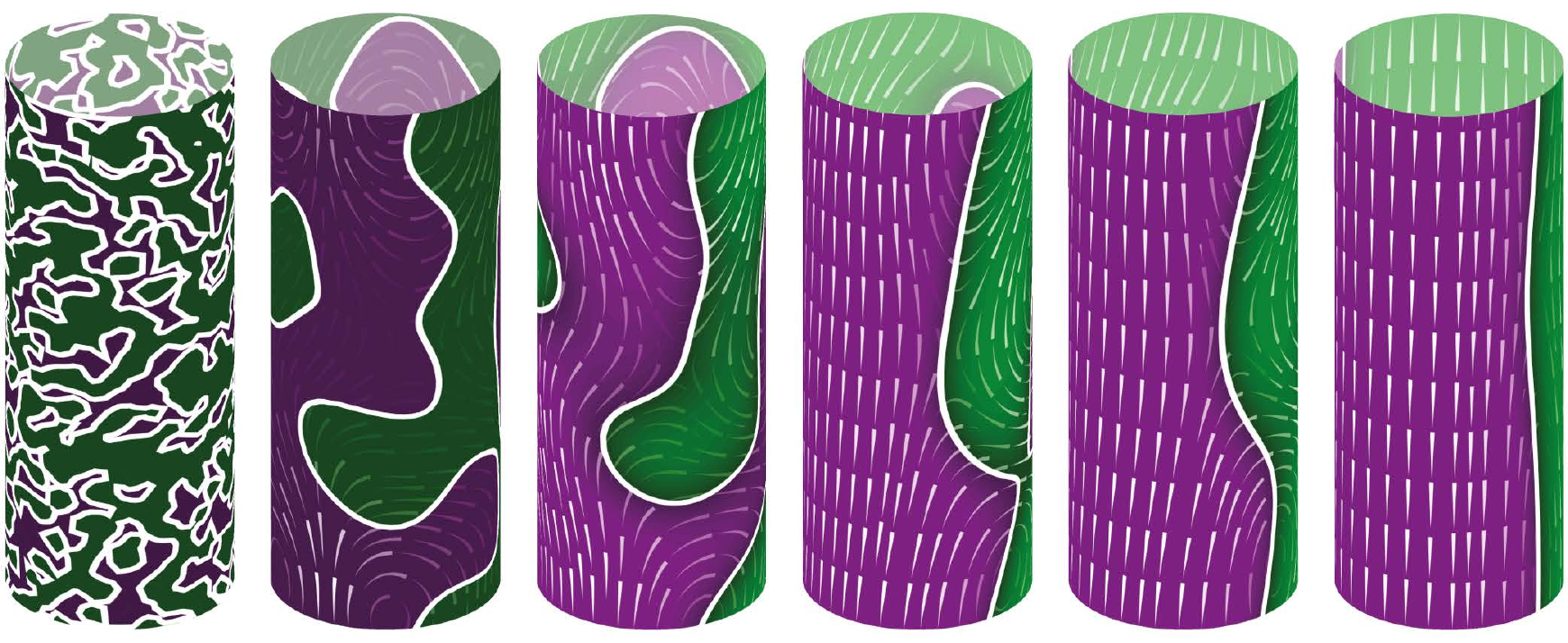}
   \caption{\label{fig:selforg} Self-organization of cytoplasmic streaming in a mathematical model of {\it Chara} \cite{WoodhousePNAS}.
Color coding corresponds to the z-component of
an order parameter associated with actin filaments at the periphery, and 
white lines represent indifferent zones separating up- and down-streaming
regions. Superimposed are streamlines of the cytoplasmic flow induced by the filament field, where the flow is directed from the thin end to the thick end
of the individual lines. Panels show progression from random disorder through local order to complete steady cyclosis.}
\end{figure*}

Two basic questions any physicist will be inclined to ask are what typical length and time scales
govern intracellular transport. This returns us to our initial question regarding the evolution of
cell size to a typical range of $1-100$ $\mu$m. Presumably mechanisms for protein targeting and
homeostatic control will be under increasing strain as the cell size increases. What, then, are the
rate-limiting factors in these processes, and how is it that cells may be able to mitigate these
limitations by internal circulation?

The internodes of characean algae present one of the most compelling test cases for these
questions. Not only are these cells among the largest in Nature, but the form of streaming found in
this system is also the most organized and symmetrical of all types of circulation currently known.
This means that the system is amenable to analysis from a geometrical point of view, and the many
physiological studies performed on this organism mean that theoretical results can be placed in the
context of detailed biological knowledge. Moreover, the fact that internodes grow from some $20$ $\mu$m
to several centimeters provides an opportunity to study mechanisms of metabolic control at a range
of scales. 

One of the aspects of streaming in the characean algae that has yet to be elucidated is just how it
affects transport of small molecules and vesicles in the cytoplasm. To the precision of current
experimental techniques, the cytoplasm appears to move as a gelled layer, retaining its shape
perhaps by way of the meshwork of ER tubes that extend throughout the endoplasmic compartment. The
temporal and spatial variations in the velocity field have yet to be quantified. Given the typical
velocities around $50$ $\mu$m/s, even the magnitude of relatively small fluctuations could be
significant, particularly when it comes to the reported caged diffusion of vesicles.

Another physical issue for which deeper understanding is needed is the precise relation between the
topology of the cytoskeletal network and the cytoplasmic flow. It would appear that streaming presents itself in increasingly organized forms as the system size is scaled up. This could be the
result of evolutionary pressure, but could also be a physical effect. So given an actin network, a
cytoplasmic rheology, and the force-velocity relations of molecular motors, can we predict in what
types of systems we will see continuous forms of circulation?

This leads us finally to comment on the possibility that some forms of streaming appear through a process of
self-organization.  As long ago as 1953 \cite{Yotsuyanagi} it was noted that cytoplasmic droplets extracted from
characean algae could spontaneously self-organize into rotating fluid bodies.  This presumably arises from
myosin motors that walk along dislodged actin filaments and entrain fluid.  Each of these motor/filament assemblies
constitutes a force dipole in the fluid, and nearby assemblies will be attracted together if nearly parallel, leading to
self-reinforcement of local order, leading eventually to long-range order.  Further evidence for self-organization comes
from much more recent work \cite{Foissner} in which steaming is completely disrupted through added chemicals
(cytochalasin and oryzalin), which, when removed, allow streaming to be reconstituted.  Strikingly, the indifferent
zone appears in a new location.  A mathematical model \cite{WoodhousePNAS} 
that combines filament bundling, flow-induced reorientation,
and coupling to the curvature of the cell wall successfully reproduces much of this phenomenology, as shown in 
Figure \ref{fig:selforg}.  

The hypothesis that streaming can self-organize through physical processes associated with hydrodynamic interactions
arising from force dipoles leads naturally to the field of `active matter', and in particular to the properties of bacterial
suspensions.  Each bacterium can be thought of as a force dipole, with one force directed backwards from the trailing 
flagella and a second directed forwards from the action of the cell body on the fluid.  The attractive interactions between
such `pusher' dipoles mentioned above has also been predicted to lead to simple large-scale flow topologies in bacterial
suspensions \cite{ConfinementPRL1}, which has subsequently been verified \cite{ConfinementPRL2,ConfinementPNAS}.  

This work was supported in part by the Leverhulme Trust, ERC Advanced Investigator Grant 247333, and the
Schlumberger Chair Fund.

%\bibliography{streaming_all_refs}

\end{document}